\documentclass[final,authoryear,5p]{elsarticle}

\usepackage{epsfig}
\usepackage{color}
\usepackage{natbib}
\usepackage{amssymb}
\usepackage{url}
\usepackage[ps2pdf,%
a4paper=true,%
breaklinks=true,%
colorlinks=true,%
pdfauthor={Tripathy et al.},%
pdftitle={}%
]{hyperref}

\journal{Advances in Space Research}

\begin{document}

\begin{frontmatter}

\title{A Study of Acoustic Halos in Active Region NOAA 11330 using Multi-Height SDO Observations}

\author[nso]{S. C. Tripathy\corref{cor}}
\ead{stripathy@nso.edu}
\ead[url]{http://www.nso.edu/staff/sushant}
\author[nso]{K. Jain}
\author[nso]{S. Kholikov}
\author[nso]{F. Hill}
\ead{kjain@nso.edu, skholikov@nso.edu, fhill@nso.edu}

\author[iiap]{S. P. Rajaguru}
\ead{rajaguru@iiap.res.in}

\author[aus]{P. S. Cally }
\ead{paul.cally@monash.edu}

\address[nso]{National Solar Observatory, 3665 Discovery Drive, Boulder, CO 80303, USA}
\address[iiap]{Indian Institute of Astrophysics, Koramangala, Bangalore 34, India }
\address[aus]{Monash Center for Astrophysics, School of Mathematical Sciences, Monash University, Clayton, Victoria 3800, Australia}
\cortext[cor]{Corresponding author}

\begin{abstract}
 
We analyze data from the {\it Helioseismic Magnetic Imager} (HMI) and the {\it Atmospheric Imaging 
Assembly} (AIA) instruments on board the {\it Solar Dynamics Observatory} (SDO) to characterize 
the spatio-temporal acoustic power distribution in active regions as a function of the height 
in the solar atmosphere. For this, we use Doppler velocity and continuum intensity observed using the magnetically  
sensitive line at 6173~\AA\ as well 
as  intensity at 1600~\AA\ and 1700~\AA. We  
focus on the power enhancements seen around AR 11330 as a function of 
wave frequency, magnetic field strength, field inclination and observation height. We find that acoustic halos occur above the acoustic cutoff frequency and extends up to 10~mHz  in HMI Doppler and AIA 1700~ \AA\ observations. Halos are also found to be strong functions of magnetic field and their inclination angle. We further calculate and examine the spatially averaged relative phases and cross-coherence spectra and find different wave characteristics at different heights. 
\end{abstract}


\end{frontmatter}

\parindent=0.5 cm

\section{Introduction}

The interpretation of acoustic waves surrounding active regions has been a difficult task since the influence of magnetic field on incident acoustic waves is not fully understood \citep[see][for a review]{moradi10}. Models point out that the waves leak into the higher atmosphere through active regions that convert a significant amount of energy into magnetosonic fast waves \citep{sch06}. These waves are then reflected back to the interior and in this process the phase of the acoustic wave is altered \citep{cally07}. It has also been argued that the helioseismic inversions for sound speed  are contaminated by surface effects associated with the strong magnetic field \citep{couvidat07}. Recent numerical simulations \citep{cally13} further suggest that processes occurring higher up in the atmosphere can significantly affect the inferences from local helioseismology in the presence of strong fields due to the contamination of the seismic signals at the surface. Two such phenomena; {\em e.g.} absorption of {\it p}-modes in sunspots around 3~mHz  \citep[][and references therein]{rjain14} and enhancement of power around 6~mHz, known as acoustic halos, are signatures of wave interaction with the magnetic field \citep[][and references therein]{sch11}. 
  
The acoustic halos were first observed in Dopplergrams as a power enhancement  around 6~mHz \citep{braun87}. This was soon followed by other studies both at photospheric and chromospheric heights \citep{braun92, brown92, toner93}. Subsequent investigations could not find such halos in continuum intensity \citep{braun92, hindman98, jain02}. Both these studies used data from the {\it Michelson Magnetic Imager} \citep[MDI,][]{sch95} on board the {\it Solar and Heliospheric Observatory} \citep[SoHO,][]{domingo95} which observes the solar photosphere in Ni~I line at 6768 \AA . The MDI Doppler signal is formed at around 125 km while the line core peaks at around 300~km \citep{fleck11, norton06}.  Enhanced power in high frequency (5.8$-$6.8~mHz) bands have also been observed in global oscillation data \citep{simoniello10}. With the availability of improved data sets mostly from space-borne instruments, both observational and theoretical studies have enriched the field and have confirmed that the chromospheric modes are the extension of the {\it p}-modes seen in the photosphere. 

 Recently \citet{sch11} inferred that the excess power is prominent for (i) moderate magnetic field strength (150 $-$ 350~G) and near horizontal ($\pm$~30$^\circ$) magnetic field, and (ii) the peak frequency increases with the field strength.  With the availability of high-cadence and high-spatial resolution full disk observations in many different wavelengths, \citet{howe12} have investigated the power, phase and coherence of the acoustic oscillations in active region NOAA AR 11072 on 23 May 2010. \citet{rajaguru13} have also examined the power maps around four active regions as a function of the magnetic field strength, inclination angle and observation height. Although there are several theories to explain the mechanism behind the halos, none of them explain all of the properties associated with acoustic halos. Recent 
forward modeling  and numerical simulations \citep{rijs15, rijs16} suggest that the power halo is wholly dependent on returning fast waves. 

In this context, our goal is to investigate  how the magnetic field influences the acoustic oscillations from the photosphere to chromosphere by examining the distribution of the acoustic power 
as a function of the height in the solar atmosphere by using multi-wavelength data on board the {\it Solar Dynamics Observatory}  \citep[SDO,][]{pesnell12}. Although this analysis is similar to that of \citet{rajaguru13}, we concentrate on the properties of a single complex active region while the former authors surveyed the average properties by combining all the four active regions. 

For the analysis presented here, we choose the NOAA active region (AR) 11330 which consist of a well defined leading spot surrounded by plage of both polarities.  We analyse the active region during 27-28 October 2011 when it was located on the central meridian at latitude 10.12$^\circ$~N (absolute heliographic longitude 249$^\circ$, see Table~1 for details).  The active region is classified as having $\beta\gamma$ configuration and is not associated with any flare during the time period considered in this analysis.   We also examine the phase and coherence spectrum between a pair of observables to study the characteristics of wave propagation. The data used in this study are downloaded from the Joint Science Operations Center (JSOC)\footnote{\url{http://jsoc.stanford.edu/ajax/lookdata.html.}} based at Stanford University. 

\section{Data Reduction} 
The multi-wavelength data, which represents different layers of the solar atmosphere, are taken from the {\it Helioseismic and Magnetic Imager}  \citep[HMI,][]{scherrer12} and the {\it Atmospheric Imaging Assembly} \citep[AIA,][]{lemen12} on board the SDO.   Specifically, we use continuum intensity ($I$) and Doppler ($V$) data  from HMI which
are produced at a cadence of 45 seconds from observations of the Fe I line at 6173 \AA\  with a spatial sampling of 0.5 arcsec/pixel which equates to a spatial resolution of 1 arcsec at optical wavelengths. The formation height of this line spans the range from 20 km above the visible surface in the wings to 270 km in the core \citep{norton06} which has been estimated using the Maltby-M umbral model \citep{maltby86} while the calculation of \citet{fleck11} shows that the mean formation height 
for the Fe I line is approximately 100~km, which is slightly lower in height than the MDI Ni I Doppler signal with a formation height of 
$\approx$ 125 km.   
\citet{fleck11} further state that the apparent formation height of the Doppler signal could increase by 40 to 50 km due to the limited spatial resolution of the instrument.  We also note that the line formation height for continuum intensity for both HMI and MDI instruments is $\approx$ 20~km \citep{norton06}.

We also use the HMI photospheric vector magnetic field measurements obtained from a second vector camera, identical to one measuring Doppler shift,  with a cadence of 135 seconds \citep{hoeksema14}.  Particularly, we use a data product designated  as the space-weather HMI active region patches \citep[SHARPS:][]{bobra14}. This series contains various parameters calculated from the photospheric vector magnetic field data and is based on the the HMI pipeline code that automatically detects active regions in photospheric line-of-sight magnetograms and intensity images and tracks the region as it rotates across the disk \citep{turmon14}. Each active region is referred to HMI Active Region Patch (HARP) and is assigned a HARP number. The SHARP data series provide maps of the photospheric vector magnetic field, inclination angle of the field to the line-of-sight, azimuth angle and the corresponding uncertainties along with many other parameters at a cadence of 720 seconds. In this study, we use the data that has been remapped from CCD coordinates to a heliographic cylindrical equal-area (CEA) projection centered on the patch and the magnetic field values are represented as spherical vector-field components. Thus vector {\bf $B$} is transformed into the components $B_r$, $B_{\theta}$, and $B_{\phi}$ and are available as FITS files. From these data sets, we compute the total magnetic field, $B$, and field inclination  angle, $\gamma$, which are average values over the period of observation used in this analysis. These  are defined as :

$$B^2  = B_r^2 +  B_{\theta}^2 + B_{\phi}^2 $$ \\ and 
$$\gamma = tan^{-1}(\frac{B_r}{\sqrt{(B_{\theta}^2 + B_{\phi}^2)}}).$$

The maximum field strength of the spot is found to be $\approx$ 3000~G and $\gamma$ ranges from $-$90$^\circ$ to 90$^\circ$ where $\gamma$ = 0$^\circ$ denotes purely horizontal field. The SHARP data is obtained from the  JSOC archive and the series hmi.sharp$_\_$cea$_\_$720s.997 corresponds to AR 11330. The remapped SHARP data has a spatial size of 917 x 536 pixels with a sampling rate of 0.03 degree per pixel.

The AIA instrument is an array of four  telescopes that provide full disk images of 4096 $\times$ 4096 pixels in different UV and EUV wavelength bands and has a spatial sampling of 0.6 arcsec/pixel with a spatial resolution of 1.5 arcsec.   For this analysis, we select two wavelength channels 1600 \AA\ and 1700 \AA\ ($I_{1600}$ and $I_{1700}$, respectively) which have been shown to be useful for helioseismic studies \citep{howe12,  rajaguru13, tripathy12}.  The 1700 \AA\ filter covers mostly UV continuum while the 1600 \AA\ filter intensity is a combination of a continuum background and some contribution of C IV. Although these observations have a cadence of 24~sec, we use non-consecutive images {\em i.e.} observations with a cadence of 48~sec which is closer to the cadence of the HMI observation. Since the HMI and AIA data have different spatial resolution and cadence, we align the images following the procedure described in section~2.1.

There are various ways to determine formation heights of intensities. \citet{uitenbroek04} calculated the optical depth unity of the continua at 1600 and 1700 \AA\ in a cross-section through a three-dimensional snapshot of solar convection. He found the heights to vary between 300 and 550 km for 1600 \AA\ and 150 and 350 km for 1700 \AA\ with a mean at about 450 and 275 km, respectively. However, as noted earlier, 1600 \AA\ filter intensity is a combination of a continuum background and some contribution of C IV. Thus, \citet{fossum05} used hydrodynamic simulations, and by folding the derived intensities with the {\it Transition Region and Coronal Explorer} 
\citep[TRACE,][]{handy99} filter transmission functions for the 1700 \AA\ and 1600 \AA\ filters 
 estimated the average formation heights to be 360 and 430 km with widths of 325 and 185~km, respectively.  Since the design of the AIA instrument is based on many features that were successful for the  TRACE instrument, we consider these derived heights to be 
the same formation heights for the corresponding passbands employed with the AIA instrument. Previously, \citet{judge01} had also estimated that the 1700 \AA\ TRACE bandpass has a typical formation height in the range $\approx$ 300$-$550~km around the temperature minimum between the upper photosphere and lower chromosphere.

However, it is important to note that most of the heights referred to here correspond to quiet Sun regions and it is highly desirable to infer the formation heights of these observables for magnetized atmospheres. There has been some evidence that in high field regions the actual formation height of the line dips below the average formation height due to the low gas pressure and density since the magnetic field also contributes to balancing the higher gas pressure of the nonmagnetic medium \citep{uitenbroek03}. For example, as described in \citet{norton06}, the formation height of the core of the 6173 \AA\ filter used in HMI instrument differs approximately by 33~km between a sunspot \citep[Maltby-M model,][]{maltby86} and quiet \citep[VAL-C model,][]{vernazza81} Sun models.   

It has been argued by \citet{howe12}, that the strong global p-mode signal in both AIA bands can not originate as high as the transition layer. It is possible that the helioseismic response in these passbands specially in AIA 1600~\AA\ band is associated with the continuum rather than the C IV line and pertains to a height range not very far above the 1700~\AA\ band.
There is also  
a possibility that a flaring event occurring higher in the transition layer could affect the power in the layers below. In order to avoid such influences, we have chosen this particular active region  where no flare occurred either in X-ray or optical wavelengths prior to, or during, the period considered in this analysis as prepared by Space Weather Prediction Center (SWPC)\footnote{ftp://ftp.swpc.noaa.gov/pub/warehouse/}. Nonetheless, given the dynamic nature of the corona, small scale and short-lived heating events (nano- and micro-flares) may occur higher in the atmosphere which may affect the brightening in the 1600 \AA\ channel. However, our method of normalizing power maps with quiet-Sun average would take care of the influence of such contributions.  Moreover, the analysis spans over a period of 16 hours and hence we believe that any contamination from transition layer as  a result of the short-lived heating events do not significantly affect the results presented here. 

\subsection{Alignment of  Data Sets}

Since the HMI images have different resolution than AIA images and can have different roll angles, we follow the guidelines described in SDO document center\footnote{https://www.lmsal.com/sdodocs} and use the SolarSoft\footnote{http://www.lmsal.com/solarsoft} routine aia\_prep.pro 
to align the HMI images to have the same resolution and orientation as the AIA images. These re-aligned images were then remapped to a spatial size of 917 x 536 pixels  using azimuthal equidistant projection  and is tracked at the Carrington rotation rate for 16 hours with a sampling rate of 0.03 degree per pixel to match the SHARP data.  We further interpolate HMI data cubes to the AIA cadence of 48 seconds.

Since the HMI and AIA images were tracked for 16 hours while the SHARP data was tracked over the life time of the AR by the HMI pipeline (the tracking rate and remapping are identical), the alignment of the images were further checked by performing cross-correlation analysis over mean images. We found that the average $B$ and $\gamma$  images were offset by 13.23 and 0.44 pixels in x and y direction, respectively compared to mean  HMI and AIA data cubes. The mean  $B$ and $\gamma$ images were appropriately shifted by the above amount so that all the images were aligned with an accuracy of  1 pixel. For the analysis described in the rest of the paper, an area of 
384 $\times$ 384 pixels covering the active region consisting of a well defined leading spot surrounded by plages of both polarity was selected in each observable. As a final step, we subtract a  running mean of 15 min from HMI and AIA data cube to remove  daily variations. Figure~1 shows context images of the active region in different observables. 

\section{Power Distribution}
Figure 2 shows the power distribution for each observable in ascending order of line  formation height (bottom to top) in four different frequency bands. The panels from left to right denote the power at 3~mHz, 5~mHz, 7~mHz and 9~mHz normalized with respect to the quiet-Sun. The quiet region patch in each observables have a size of 384 x 384 pixels in each spatial direction, and was placed at the same latitude and longitude as the AR. This data was processed in an identical way as the active region i.e. remapped using azimuthal equidistant projection and tracked at the Carrington rate for 16 hours on 09 June 2010. The start and end time of each observable is tabulated in Table~2.  The power maps are produced by calculating the power in each pixel of the data cubes using a fast Fourier Transform (FFT) and  summed over a 1~mHz band of frequencies centered every 0.1~mHz in the frequency range of 1~$-$~10.5 mHz.  The maps were further smoothed over three bins in each spatial direction so that any fractional misalignment between different observables do not bias the result presented here.    

As is known from earlier studies involving MDI data \citep{hindman98, jain02}, the power maps of continuum  intensity observation (bottom panels) do not show any  power halos since the line formation height of the intensity observable for both MDI and HMI instruments is similar ($\approx$ 20~km) and is well below the formation height of HMI $V$ and AIA intensities. From the Doppler observations, we notice the well-known 
suppression of power in the 3~mHz band \citep{woods81, lites82, brown92}, and enhancement of power at higher frequencies which is manifested as enhanced power rings around the sunspot at 7~mHz. At 9~mHz, we observe more diffused power enhancement over the active region. The narrow regions of power enhancement inside the umbra of the sunspot in the 7~mHz and 9~mHz band appears to be an artifact. These could be due to the leaking of
magnetic field to the Doppler velocity observation near the strongest field strength since the Fe I spectral line at 6173~\AA\ 
is magnetically sensitive. \citet{hoeksema14} found that the velocity sensitivity of the daily variation increases with field strength.  \citet{couvidat16} further showed that the line-of-sight algorithm produces significant errors in the presence of strong magnetic field as the shape of the Fe~I line in a strong and inclined field differs significantly from the synthetic profile and  Voigt profile used to produce the look-up tables in places where high velocity is present \citep[see Fig. 22 of ][]{couvidat16}   

The power maps for AIA 1700 \AA\ and AIA 1600 \AA\ show more complex structures. The network regions clearly show the 5 min acoustic oscillations. At the higher frequency of 5~mHz, the power is suppressed in both AIA channels while at 7~mHz and 9~mHz excess power around the sunspot is seen in AIA 1700 \AA\  similar to the Doppler observations. However, AIA 1600 \AA\ observations show the presence of {weak enhancements at very few locations as compared to 1700 \AA\ channel. This can be more clearly visualized in Figure~3, where we have superposed the contours  on the power maps. The colors blue, yellow and red represent enhanced power levels of 1.2, 1.5, and 2.0 over the quiet Sun,  respectively and distinctly illustrates the power halos for HMI $V$ and AIA 1700~\AA\ observables.} For Doppler observations, our analysis is consistent with the study of  \citet{howe12}  where enhancements were seen as rings around the sunspot for AR 11072. However differences are seen in the  AIA channels. \citet{howe12}  find high-frequency halos around the entire active region while we observe partial and diffused power halos. In our previous study \citep{tripathy12} full halos were observed around AR 11092. Since both AR 11072 and 11092 were simple active regions while AR 11330 was more complex and covers the entire cube (See Figure~1), we presume that it is the complexity of the active region that controls the extent of the halos at greater height.  This assumption appears to be supported through the analysis of \citet{rajaguru13} where four active regions consisting of two simple and two moderately complex ARs were analyzed.  For simple ARs, full halos were observed at AIA wavelengths while for complex or extended active regions, the halos appeared to be diffused. It is also to be noted that 
observations made in the same AIA band by the  TRACE instrument reveal contradicting results. While \citet{krijger01} 
reported halos or ``aureoles" in both AIA 1700 \AA\ and AIA 1600 \AA\ channels, \citet{muglach03} reported 
power decrease in the surroundings of the active region when the frequency was binned over 5.5$-$7.5~mHz. The later results could be due to the adopted analysis procedure where all power with a probability less than 95\% was set to 0 and not included when calculating the frequency bins.  Since the underlying mechanism of power enhancement is still being debated and appears to vary between different active regions, a statistical investigation of power maps of many active regions is warranted but is beyond the scope of the present paper.      

\subsection{Magnetic Field and Power Distribution}
One of the major advantage of SDO data is the availability of vector magnetic field measurements. In this study, we use the magnetic field strength, $B$, and the field inclination, $\gamma$,
to analyze the power distribution as a function of the frequency. Since, we are interested to investigate the power enhancements and not suppression which occurs in the umbra and penumbra regions of the sunspot, we  restrict 
our analysis to regions where $B$ is less than 850~G to be consistent with the study of \citet{rajaguru13}. The normalized power {\em i.e.} the power of the active region normalized by the power of the quiet region is averaged over 10~G bins in $B$ and 4$^\circ$ in $\gamma$ so that the power distribution can be analyzed either as a function of the magnetic field or field inclination or a combination of the two. The choice of grouping is identical to that used in \citet{rajaguru13}.

Figure~4 shows the power as a function of the magnetic field and frequency, where the magnetic field is averaged over 10~G bins. The panels show the power structure for the Doppler velocity and two AIA intensities. The first thing to notice is that the 
power enhancement begins at about 5~mhz for $V$ and 6~mhz for $I_{1700}$ and extends up to 10~mHz in both observations. 
At the lowest height corresponding to the Doppler velocity (neglecting the continuum intensity since no halos are seen), the halos can be observed  both at weak and strong field regions and the structure varies with frequency. We also notice power reduction for very low magnetic field between 6$-$8~mHz range. As we progress to greater heights, we observe that the halos are confined to intermediate 
field strengths ($B \approx$ 350~G) at the height of $I_{1700}$,  while at the height of $I_{1600}$, we detect very localized enhancements around$B \approx$ 700$-$800~G which are not at the locations of the halo and appear to correspond to 
the power enhancements seen near the sunspot boundary in Figure~2.

In contrast, the simulated power structure presented in \citet{rijs16} shows different characteristics as compared to our observations presented here.  The simulated Doppler observations show formation of halos over weak to moderate field (50~G $<$ B $<$ 700~G) and can be seen up to 10~mHz for low $B$ with a band of power reduction at higher frequencies. However, the power distribution for the synthetic AIA bands shows power excess only at low values of {$B <$ 100~G} which is very different from our measurements.  The difference may be attributed to the topology of the analyzed active region as compared  to the numerical simulations with a simplified axisymmetric magnetic field and field inclination. 

We next explore the power as a function of the field inclination angle averaged over 4$^\circ$ bins (Figure~5).  For Doppler observations, power halos begin at about $\nu > 6$~mHz but are mostly confined to low to moderate and positive inclined fields  ($\gamma \approx 0^\circ - 45^\circ$). As the field nears vertical orientation, the halos shift to higher frequencies, and for vertical fields the halos are confined to frequencies of about 9.5~mHz. When the inclination is negative (field pointing in the opposite direction) power enhancement is only observed at frequencies $\gtrsim$ 8~mHz. The power distribution, however,  is very different for both $I_{1700}$ and $I_{1600}$. For $I_{1700}$, the power enhancement is weak and confined to moderate to purely vertical field, while for $I_{1600}$, the enhancement is seen only when the field is nearly horizontal.   However, it should be noted that the $B$ and $\gamma$ values that we have used in this study are photospheric values and do not represent actual values at AIA observation heights. In the future, we plan to use observations from Synoptic Optical Long-term Investigations of the Sun (SOLIS) instrument which would yield vector magnetic field observations at chromospheric heights.  As demonstrated in \citet
{bloomfield07} vector field measurements at different heights will also aid to spatially align the pixels between different observables. 

We further scrutinize the power distribution as a function of $B$ and $\gamma$ by sub-dividing each of them into different groups based on the analysis of  \citet{rajaguru13}; (i)   three different range of inclination angle as a function of the magnetic field:  nearly horizontal field 
($|\gamma|~<~16^\circ$), moderately inclined fields (16$^\circ~<~|\gamma|~<60^\circ$), and nearly vertical fields ($|\gamma|~>~60^\circ$)  and (ii) as a function of $\gamma$ for three different ranges of $B$ {\em viz.}  $B < 100$~G,  
100~G$~ < B < 200$~G, and  200~G $< B < 450$~G.  These distributions as a function of the atmospheric height is illustrated in Figures~6$-$7. 
A detailed examination of Figures~6$-$7 reveal how the power distribution of acoustic modes are modified by the combined $B$ and $\gamma$ values. In each of these figures, panels from bottom to top represent different heights in the solar atmosphere in ascending order.    
The power distribution corresponding to the Doppler signal, which is illustrated in the lower panels, we identify strong halos for $\nu$ $\gtrsim$ 5~mHz up to about 10~mHz except for a narrow region between $ 6<\nu < 8$~mHz for  $B \lesssim$  100~G. This band of reduced power between two bands of power excess is clearly visible in the left bottom panel of Figure~7 where the power distribution is shown as a function of $\gamma$. The reduced power band is further noticed for highly inclined field (right panel of Figure~6). We also observe that, at low frequencies, strong halos occur only for horizontal and moderately inclined field. As the field becomes more vertical, the halos shift to higher frequencies.

Power distribution from greater heights corresponding to AIA 1700~\AA\ and AIA 1600~\AA\ bands (middle and top panels of 
Figures~6$-$7 indicate that the pattern of power halos have changed with height. At about 385~km where the AIA 1700~\AA\ line forms,
excess power is seen at low to moderately $B$ values starting at about 6~mHz. As for Doppler observations, halos can be seen up to 10~mHz. We also notice that halos are absent when the field is horizontal and as a result the power halos appear 
as two disjoint columns in $I_{1700}$ and $I_{1600}$ maps. This power distribution was also observed  by \citet{rajaguru13} and has been termed ``twin halos". As $B$ increases, the twin halos shift to higher $\nu$ and the separation between them increases implying that halos at higher magnetic field exist only when the field is vertical. In the case of $I_{1600}$, the twin halos are observed for low magnetic field (left top panel of Figure~7). 

The most accepted mechanism of  halo formation relies on the attack angle {\em i.e.} the angle between the wave vector of the incident wave and the orientation of the magnetic field at the conversion/transmission depth \citep{sch06}.  If the attack angle is large, energy could be transported from the fast-acoustic mode to the fast-magnetic mode while for small attack angles, the energy will be mainly channeled  into the field aligned slow mode. Thus for Doppler observations near disk center, the attack angle is large between the velocity and the horizontal magnetic field and produces strong halos for low to moderate magnetic field only when the field is horizontal concurring with the observations of \citet{sch11}. This theory also explains the spreading of the halos with height; since the magnetic canopy where the sound speed equals the Alfv\'{e}n velocity is located at greater heights due to the spreading of the magnetic field.  

\section{Phase and Coherence Spectrum}  
The phase difference and cross-coherence function between a pair of observables (e.g. 1 and 2) are defined as  

\begin{equation}
    \label{eq:1}
{\delta\Phi_{12}(\nu)= arg (P_1(\nu)\; .\; P_2^*(\nu)),\\
}
\end{equation}
where $P_1(\nu)$ and $P_2^*(\nu)$ are the Fourier transforms of the power spectra of a pair of observables; $*$ denote the complex conjugate and $\nu$ is the temporal frequency. We adopt the convention that positive values of $\delta\Phi_{12}$ imply that signal 1 leads signal 2 {\em i.e.} the wave is propagating from height 1 to 2 and the opposite holds for a negative phase shift. Here we only consider the spatially averaged phase difference given per sampled frequency by 

\begin{equation}\label{eq:1b}
{[\delta\Phi_{12}]_{xy}(\nu)= arg ([P_1(\nu)]_{xy}\; . \; [P_2^*(\nu)]_{xy}),\\
}
\end{equation}
where the square bracket denotes averaging over the spatial direction \citep{lites98, krijger01} . The degree of coherence between the two signals is expressed as 
\begin{equation}
    \label{eq:2}
{C_{12}(\nu)= \frac{|<(P_1(\nu).\, P_2^*(\nu)>|}{\sqrt{<|P_1(\nu)|^2> <|P_2(\nu)|^2>}},
}
\end{equation}
where the expectation value is approximated as a running mean in the frequency interval \citep{lites98} because without any smoothing the coherence between two sinusoidal Fourier components at given spatial and temporal point is unity regardless of the corresponding Fourier amplitude and phase difference. For pure noise this procedure yields a positive coherence  $C = 1/\sqrt{n}$ where $n$ is the frequency bin used in the averaging procedure.    In this analysis we consider N = 21 which yields a frequency smoothing over 0.357~mHz and a pure noise, $C_{12} = 0.22$. 

Figure~\ref{figure8} shows the spatially averaged phase difference and coherence spectra between Doppler and other observables.  The spatially averaged phase difference and coherence spectra as a function of the frequency is plotted in one figure so that one can correlate the behavior between the two.  For comparison,  we also include the phase difference and coherence spectra between the two AIA channels. 

The figure illustrates that the coherence is larger between a pair of  quiet regions (blue lines) compared to the active regions (red lines). Since the active region changes the properties of the waves either through mode conversion or through scattering, transmission and absorption (or a combination of these phenomena), it is expected that the coherency between the two layers should decrease. The quiet Sun phase difference between $V$-$I$ (panel a)  has a value of about 135$^\circ$ in the 5~min band and agrees with the earlier findings \citep{jimenez90}. We also notice a difference between the quiet and active regions; for the quiet regions we note a positive phase lag  at all frequencies, while we find a negative phase lag at about 6~mHz for the active region implying a reflected or refracted wave at higher frequencies. Moreover, the phase difference for quiet (active) region slowly decreases (increases) to 0$^\circ$ close to the Nyquist frequency.  The coherence between  $V$-$I$ is high at low frequencies but slowly decreases. Beyond 2.5~mHz, the coherence again increases and attains a peak value at $\lesssim$ 4~mHz before decreasing to the noise level at high frequency of 10~mHz. It is also interesting to note that the pattern of phase shifts and coherence beyond ~6.5~mHz is similar  to the phase difference pattern of the active region.     
 
The phase difference between HMI $V$ and $AIA$ intensities show a sharp peak about 2~mHz, around the upper limit of the granulation signal. The phase difference reverses its sign about 3~mHz implying a downward propagation and reaches $-180^\circ$ close to 10~mHz.  As discussed earlier,  the coherence decreases for the active region. Although the coherence spectra between $V$-$I_{1700}$ and  $V$-$I_{1600}$
are similar, we note differences between the phase information in both quiet and active regions. The phase difference between $V$-$I_{1600}$ corresponding to the quiet region  implies a upward propagating wave while for $V$-$I_{1700}$, we note an evanescent wave at higher frequencies. Moreover the phase difference between $V$-$I_{1600}$ stabilizes about $-90^\circ$ implying a transverse wave at the highest frequency. \citet{rijs16} have also compared the phase shifts between the two AIA intensities with respect to the continuum intensity as a function of the magnetic field strength.  The simulation shows positive phase differences at higher frequencies. 

\citet{howe12} have also constructed relative phase and cross-coherence maps (not spatial averages) between different observables  and found that their properties are altered around the active region. Maps of phase-shifts between $I$ and $I_{1700}$ have also been investigated by \citet{rajaguru13a}. However, a direct comparison between these results is not feasible since we have focused on spatial averages instead of maps of the entire active region.

\subsection{Comparison between AIA 1700~\AA\  and 1600~\AA\ }

The phase difference between $I_{1700}$ and $I_{1600}$ follows the expected behavior of propagating waves up to about 8~mHz after which it decreases slowly.  The same behavior was earlier seen by \citet{krijger01} where the decrease of the phase difference was attributed to acoustic waves steepening into weak shocks on the way up. This is, however, not supported in the wave simulation performed by \citet{fossum05}. The quiet Sun signals are coherent up to about 8~mHz and gradually decrease to lower values. The coherence between the active patches shows a similar behavior with a sightly smaller magnitude. At higher frequencies, where the coherency decreases, the active region has a marginally higher coherence than the quiet region. The turnover interestingly occurs at the frequency where the phase difference also begins to decline. 

\section{Summary}
We have used data from HMI and AIA on board SDO to understand the behavior of acoustic oscillations in the photosphere and low chromosphere. The major motivation of this work is to examine the spatio-temporal power distribution as a function of the magnetic field strength, inclination of the field and height in the solar atmosphere. We also focused on the coherence and phase difference properties between different observables. 

We find that the strong halos occur at frequencies above the acoustic cutoff extending up to 10~mHz in both Doppler and AIA 1700\AA\  channel for low and moderate magnetic field strength and horizontal field providing evidence that mode conversion is responsible for the formation of the halos.  It is also observed that the halos are strong functions of  magnetic field and field inclination.  We also note the spreading of the halos with height since the magnetic field fans out at greater heights increasing the height where the sound speed equals the Alfv\'{e}n velocity. 
Although, we find some agreement between the halos formed in numerical simulation and Doppler observation, we find large discrepancy between AIA observation and synthetic AIA data obtained from the simulation primarily due to complex nature of the observed active region.  We also observe that the differences in phase between HMI V and AIA intensities  
are negative beyond 5 or 6~mHz reassuring that the reflection or refraction of the wave occurs at a certain height and as a result the wave propagates  downward.    We, however, note the $B$ and $\gamma$ values that we have used in this study are photospheric values and do not represent actual values at AIA line formation heights.  In future, we plan to use field extrapolation to the observing height or simultaneous vector field measurements at different heights for a robust conclusion.  
 \section*{Acknowledgments}

{\it SDO} data courtesy of SDO (NASA) and the HMI and AIA consortium. This work was partially supported
by NASA grant NNH12AT11I to the National Solar Observatory. This research has made use of NASA's Astrophysics Data System and was performed
under the auspices of the SPACEINN Framework of the European Union (EU FP7). We thank John Leibacher for useful comments on the manuscript. We also thank the reviewers for their critical comments.  

\bibliography{/home/sushant/paper/aastex52/sct.bib}

\onecolumn

   \begin{table}
\caption{Properties of the Data Sets in Different Observables. The start and end time correspond to the T\_OBS keyword in the image file. For HMI, the keyword reflects the peak of the weighting function applied to the constituent images comprising the sequence. For AIA, it is the center of the time for which the camera shutter is open. The unit of time for HMI data is International Atomic  Time (TAI) while the unit for AIA data is Coordinated Universal Time (UTC). See \url{https://www.lmsal.com/sdodocs} for more information.}
\begin{tabular}{lcclcc}
\hline
Instrument/&Spatial &Cadence&Line Formation &Start time&End time\\
Observable&Sampling (''/pixel)&(sec)&Height (km)&& \\
\hline
HMI/Doppler & 0.5&45&100 $\pm$ 50 &2011.10.27 21:50:12 &2011.10.28 13:50:12 \\
HMI/Continuum Intensity & 0.5&45&20  &2011.10.27 21:50:12 &2011.10.28 13:50:12 \\
AIA/1700 \AA\ &0.6&48&430 $\pm$ 185&2011.10.27 21:50:32 &2011.10.28 13:50:32 \\  
AIA/1600 \AA\ &0.6&48&360 $\pm$ 325& 2011.10.27 21:50:18 &2011.10.28 13:50:18\\
\hline
\end{tabular}
\label{table1}
\end{table}

\begin{table}
\begin{center}
\caption{Start and End time of Quiet Region in Different Observables.}
\begin{tabular}{lccc}
\hline
Instrument/&Start time&End time\\
Observable&& \\
\hline
HMI/Doppler   &2010.06.09 19:07:37 &2010.06.10 11:07:38 \\
HMI/Continuum Intensity  & 2010.06.09 19:07:37 &2010.06.10 11:07:38 \\
AIA/1700 \AA\ &2010.06.09 19:08:08 &2010.06.10 11:08:08 \\
AIA/1600 \AA\ &2010.06.09 19:07:54 &2010.06.10 11:07:54 \\
\hline
\end{tabular}
\end{center}
\label{table2}
\end{table}
\begin{figure}[ht]
\begin{center}
\includegraphics*[width=14cm]{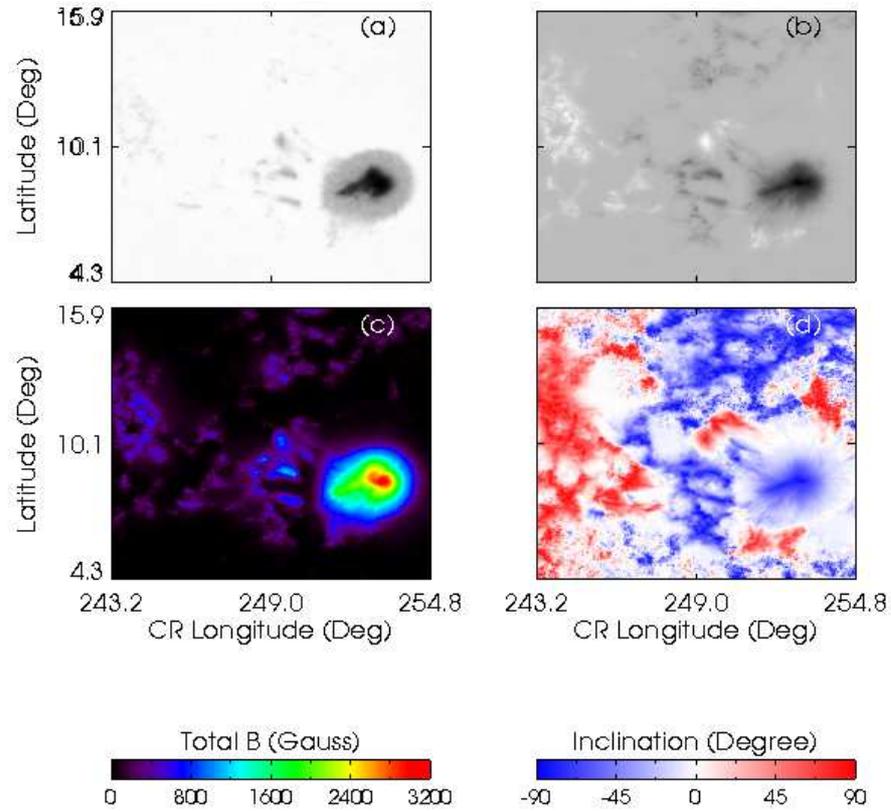}
\end{center}
\caption{Mean images of AR NOAA 11330 in different HMI observables; (a) continuum intensity, (b) line-of-sight magnetogram, (c) total magnetic field strength, and  (d) the magnetic field inclination. The last two quantities are derived from the vector magnetic field maps as described in the text. \label{figure1}}
\end{figure}

\begin{figure}[ht]
\begin{center}
\includegraphics*[width=14cm]{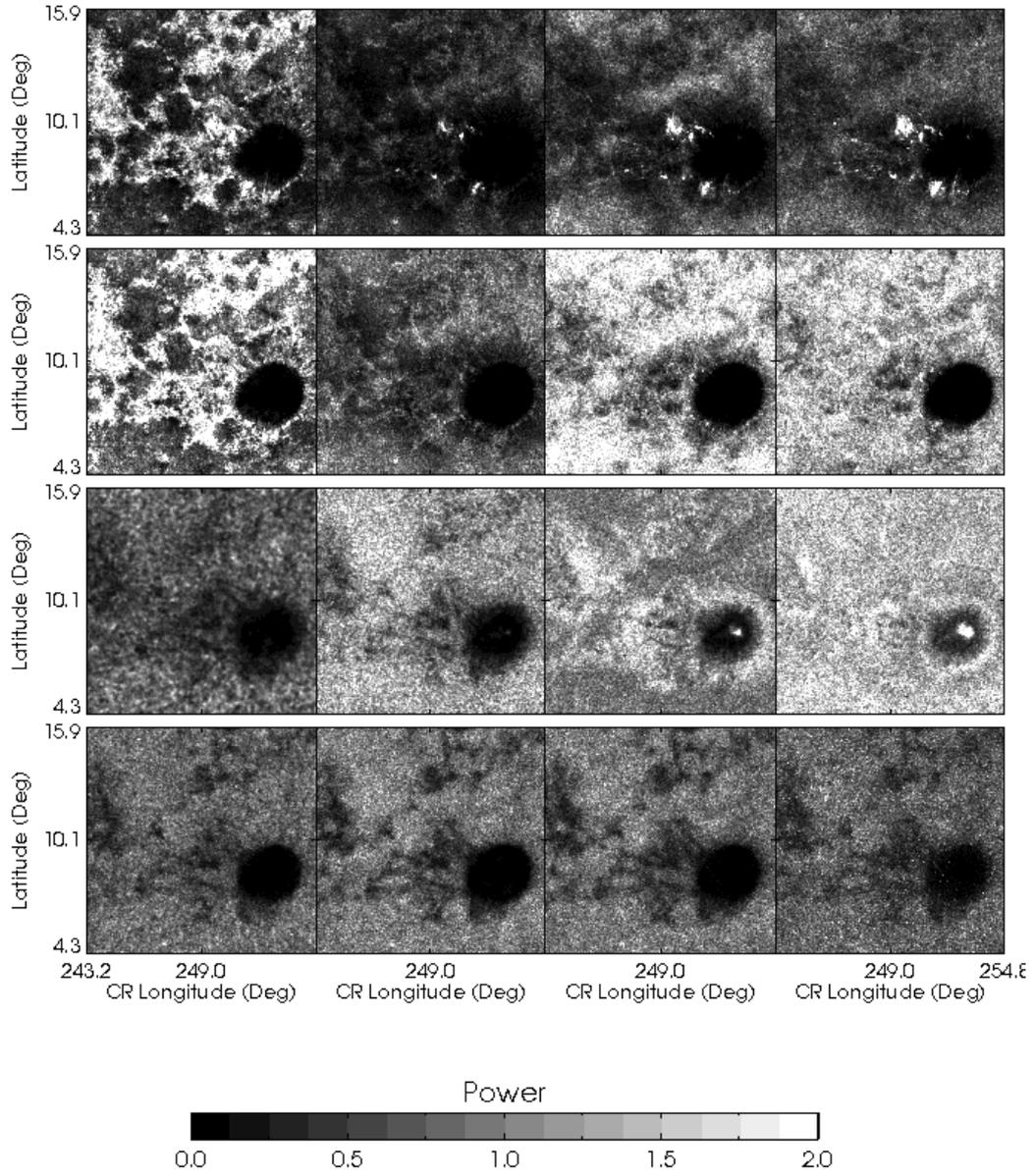}
\end{center}
\caption{Power of the active region at four representative frequencies and four different observation heights. The panels from left to right denote the power normalized with respect to the quiet-Sun at 3~mHz, 5~mHz, 7~mHz and 9~mHz. The rows from bottom to top represent heights in the increasing order; HMI continuum intensity, ($I$), HMI velocity ($V$), AIA 1700 \AA\ ($I_{1700}$) and AIA 1600 \AA\ ($I_{1600}$) channels. 
\label{figure2}}
\end{figure}

\begin{figure}[ht]
\begin{center}
\includegraphics*[width=14cm]{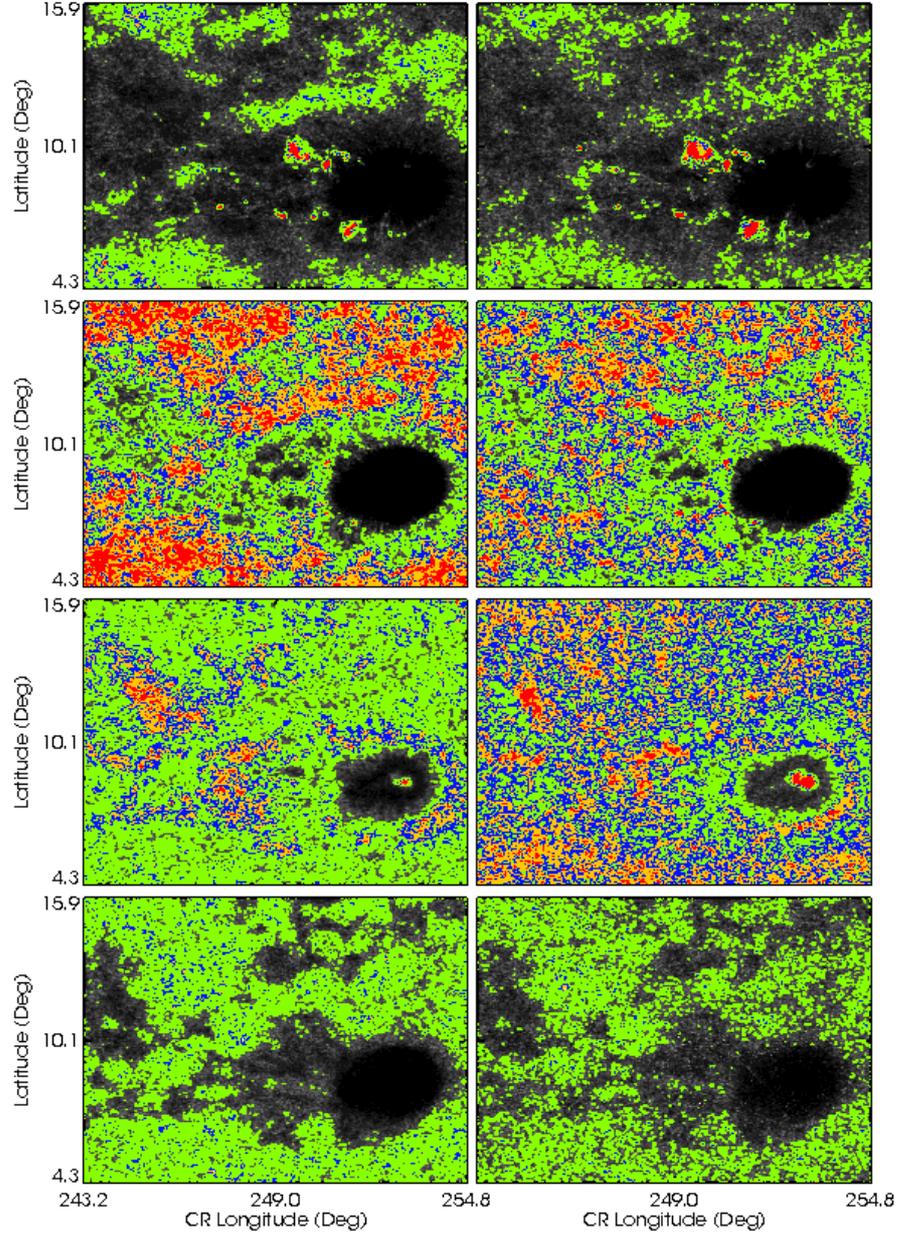}
\end{center}
\caption{Power maps at  5~mHz (left panels) and 7~mHz (right panels) bands shown with filled and colored contours. The colors represent different power levels with respect to the quiet Sun; green:0.8, blue: 1.2, yellow: 1.5, and red: 2.0. The rows from bottom to top represent heights in the increasing order; HMI continuum intensity, ($I$), HMI velocity ($V$), AIA 1700 \AA\ ($I_{1700}$) and AIA 1600 \AA\ ($I_{1600}$) channels. \label{figure3}}
\end{figure}

\begin{figure}[th]

\begin{center}
\includegraphics*[width=\textwidth]{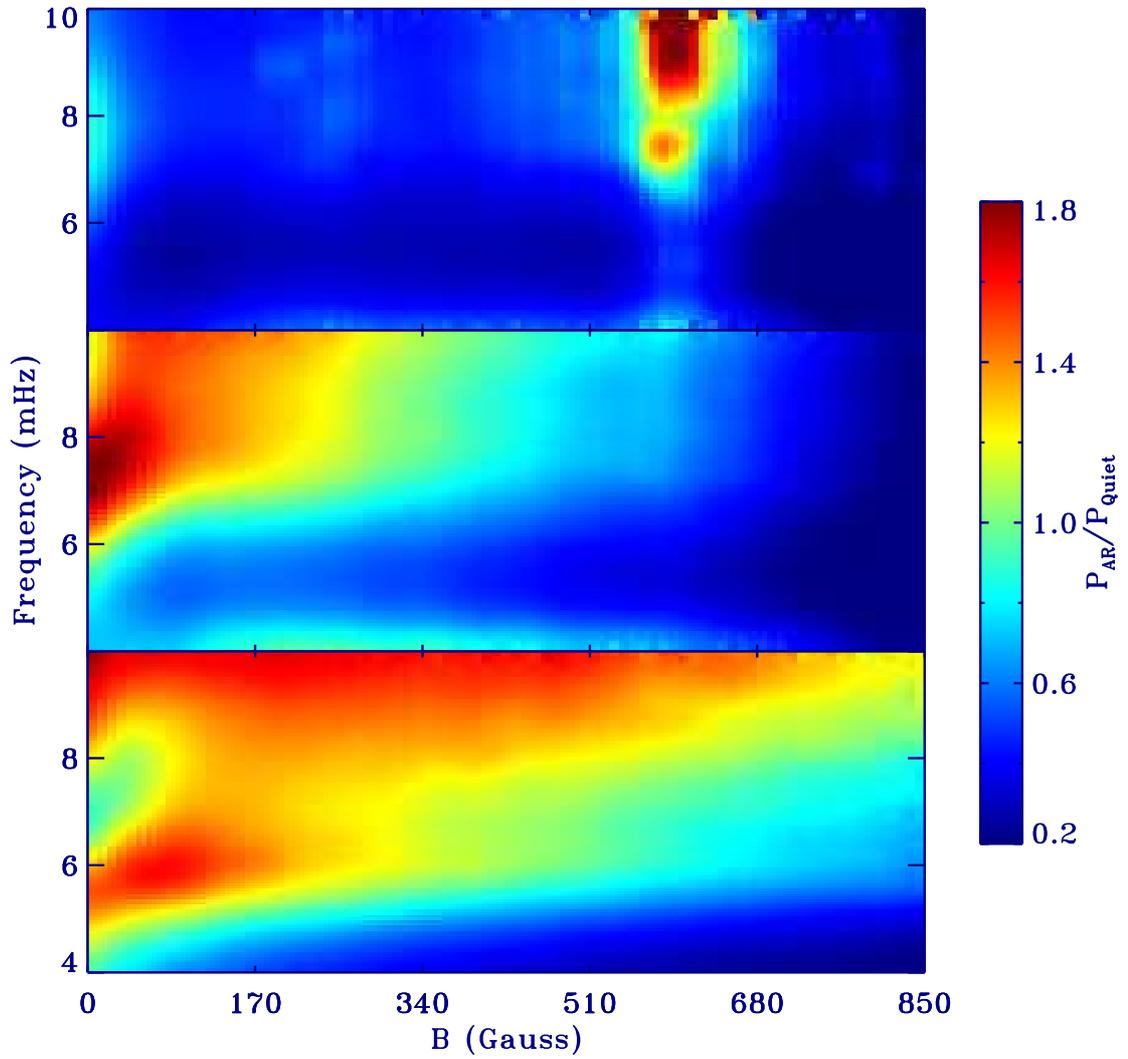}
\end{center}
\caption{Normalized power as a function of total magnetic field strength, $B$, binned over 10~G with a cutoff value of 850~G. Panels from bottom to top represent different heights in the atmosphere; (bottom) HMI V, (middle) AIA 1700~\AA , and (top) AIA 1600~\AA . \label{figure4}}
\end{figure}
\begin{figure}[th]

\begin{center}
\includegraphics*[width=\textwidth]{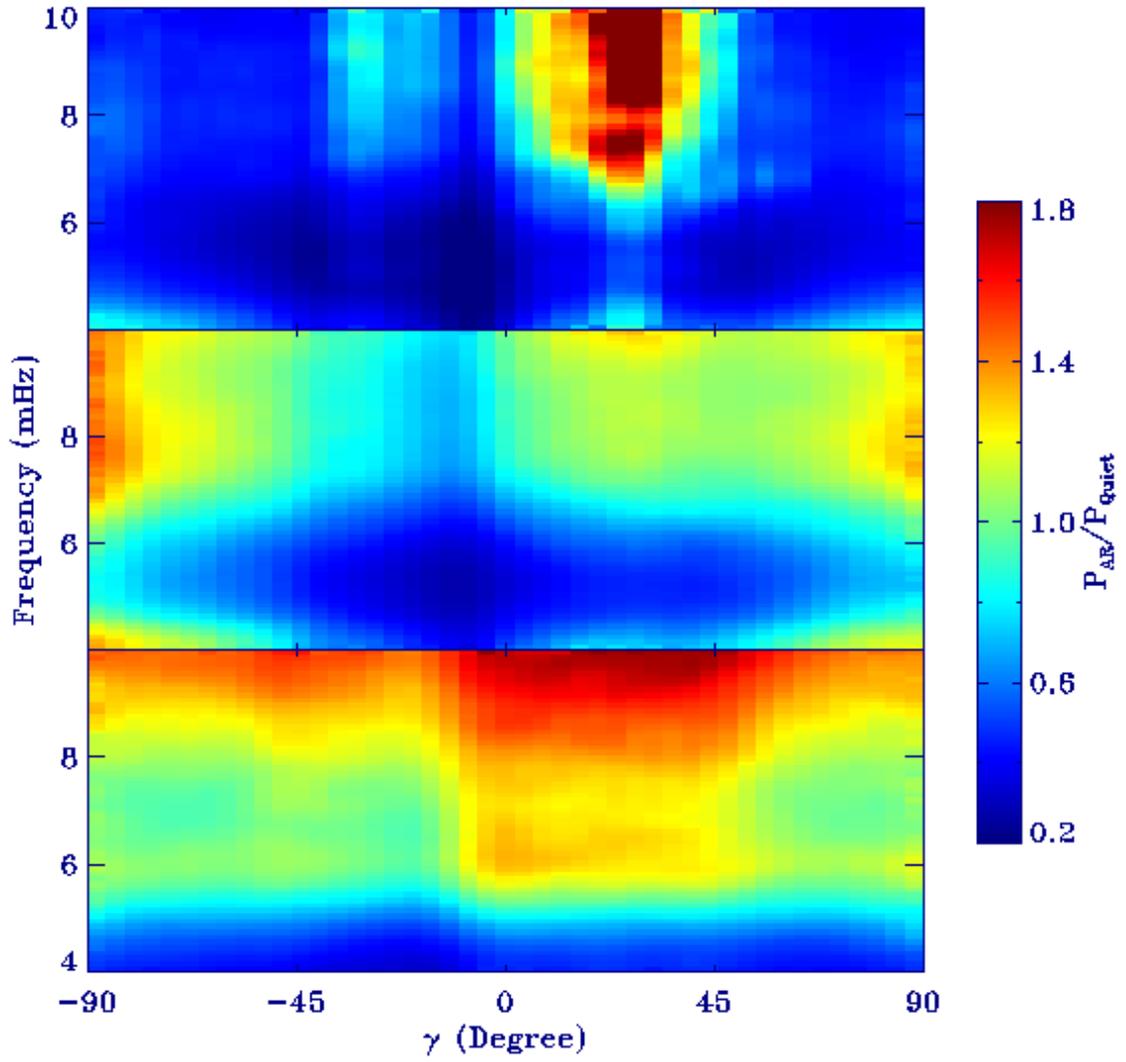}
\end{center}
\caption{Normalized power as a function of the magnetic field inclination, $\gamma$, averaged over $4^\circ$ bins. Panels from bottom to top represent different heights in the atmosphere; (bottom) HMI V, (middle) AIA 1700~\AA , and (top) AIA 1600~\AA .\label{figure5}}
\end{figure}
\begin{figure}[h]
\begin{center}
\includegraphics*[width=\textwidth]{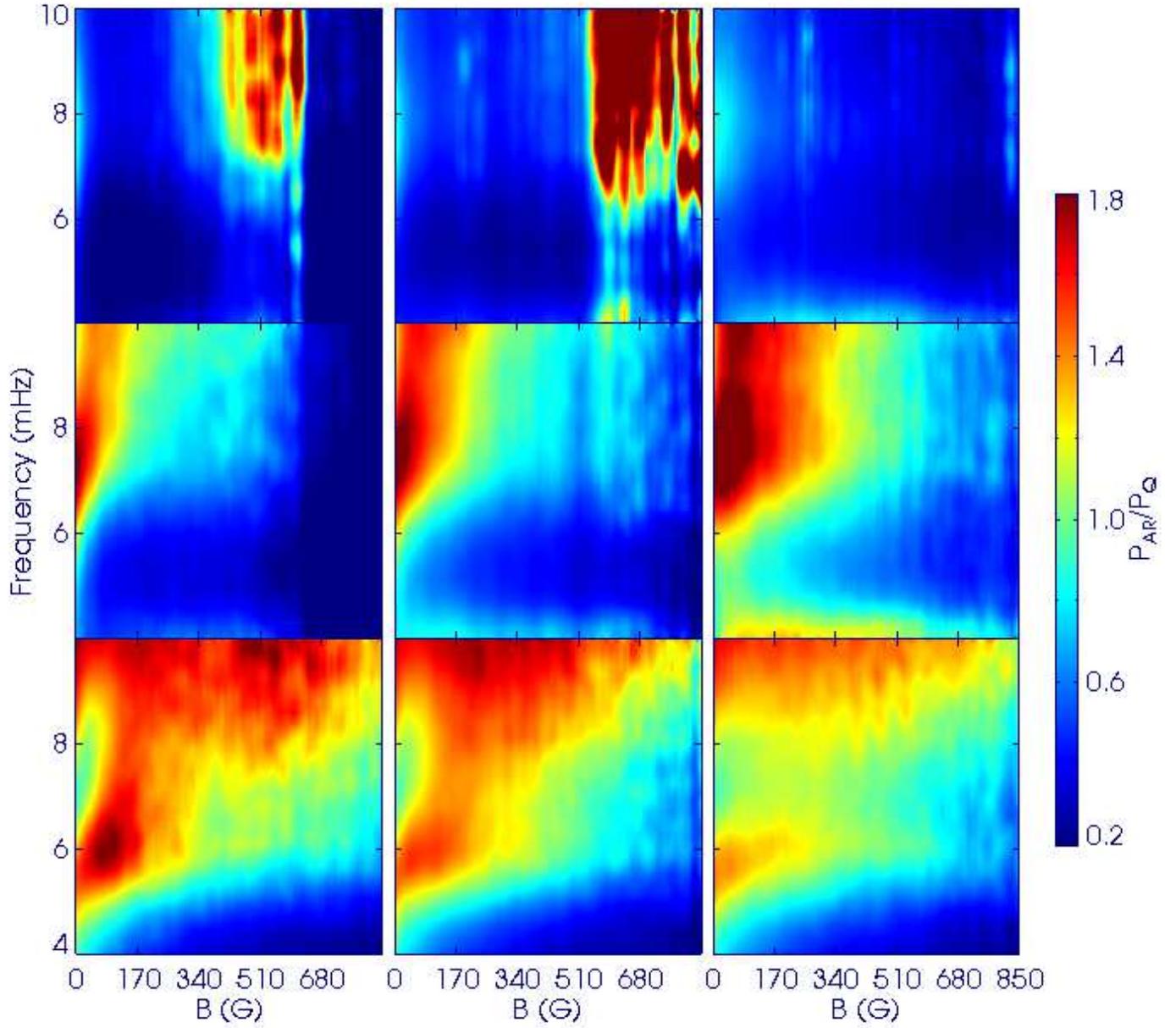}
\end{center}
\caption{Power as a function of total magnetic field strength, $B$, averaged over three different field inclination angle: (left) nearly horizontal field ($|\gamma| < 16^\circ$), (middle) inclined fields ($ 16^\circ~<~|\gamma|~< 60^\circ$), and (right)  nearly vertical fields ($|\gamma|~>~60^\circ$). Panels from bottom to top represent different heights in the atmosphere; (bottom) HMI V, (middle) AIA 1700~\AA , and (top) AIA 1600~\AA .\label{figure6}}
\end{figure}
\begin{figure}[h]
\begin{center}
\includegraphics*[width=\textwidth]{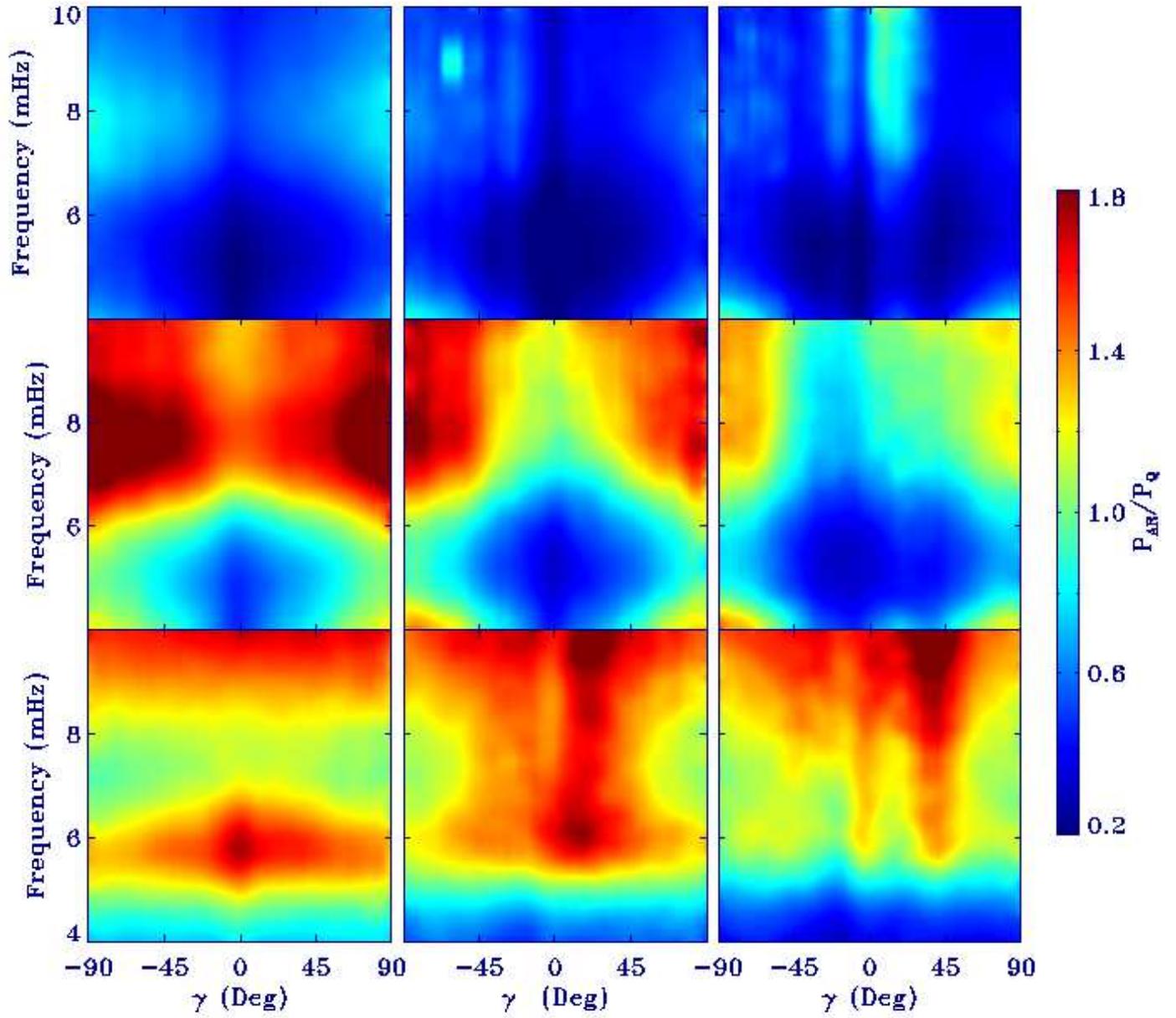}
\end{center}
\caption{ Power  as a function of $\gamma$ averaged over different ranges of total magnetic field strength, $B$: (left) $B < $100~G, (middle) $100 < B < 200$ G, and (right) $200~G <  B < 450$ G.   Panels from bottom to top represent different heights in the atmosphere: (bottom) HMI V power, (middle) AIA 1700~\AA , and (top) AIA 1600~\AA .\label{figure7}
}
\end{figure}

\begin{figure}[ht]
\begin{center}
\includegraphics*[width=\textwidth]{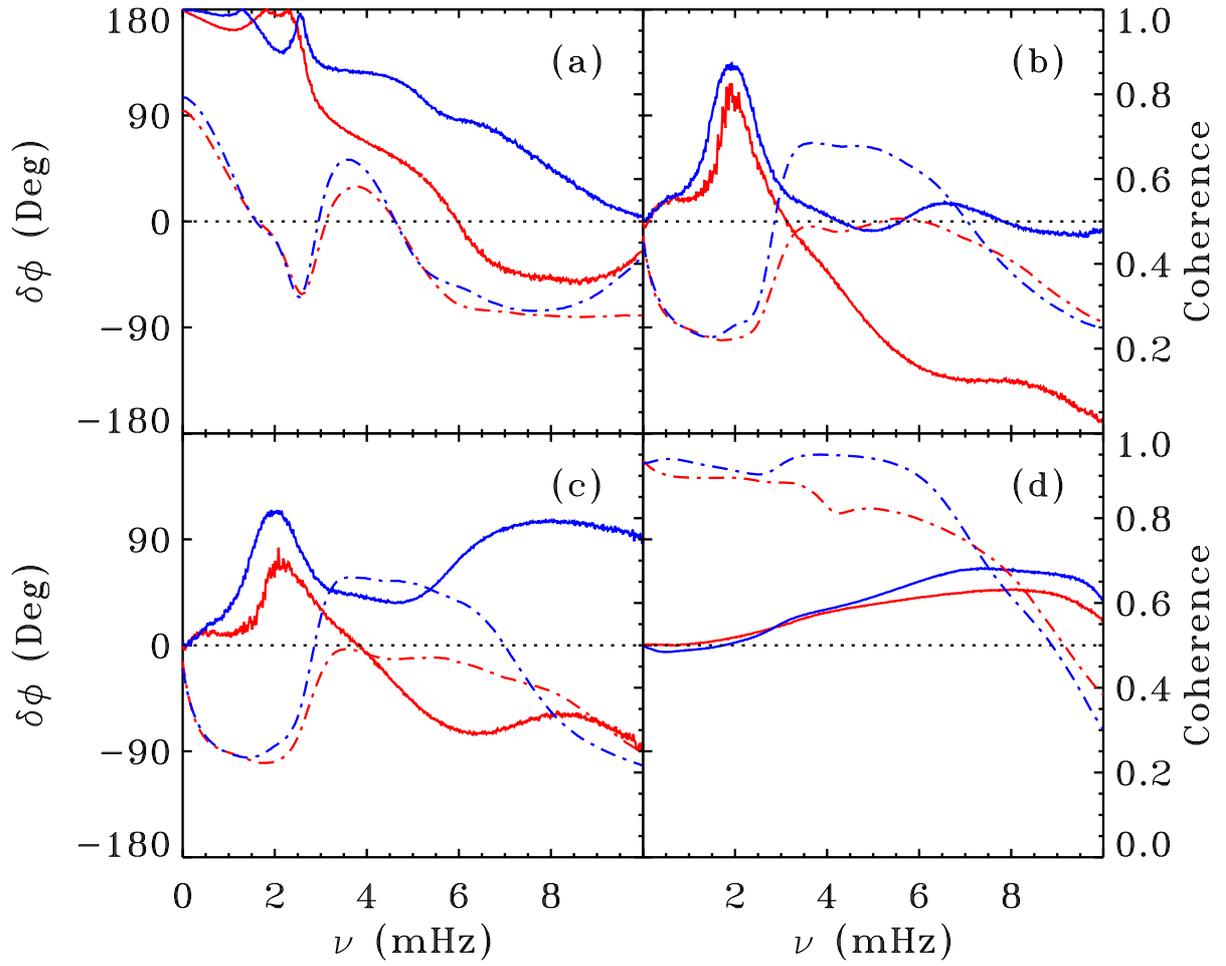}
\end{center}
\caption{Spatially averaged phase difference (solid line) and coherence spectrum (dash-dotted line) between two different data sets.  The blue and red colors represent quiet and active regions, respectively. The scale on the left corresponds to the phase difference spectra while the coherence spectra has its scale on the right. Panels denote the parameters between (a) $V$ and $I$, (b) $V$ and $I_{1700}$ (c) $V$ and $I_{1600}$, and (d) $I_{1700}$ and $I_{1600}$. \label{figure8}}
\end{figure}


\end{document}